# MBE grown Self-Powered β-Ga$_2$O$_3$ MSM Deep-UV Photodetector

Anamika Singh Pratiyush[1,a)], Sriram Krishnamoorthy[2,3], Sandeep Kumar[1], Zhanbo Xia[2], Rangarajan Muralidharan[1], Siddharth Rajan[2], Digbijoy N. Nath[1,a)]

[1]Centre for Nano Science and Engineering (CeNSE),
Indian Institute of Science (IISc), Bangalore 560012
[2]Department of Electrical and Computer Engineering, The Ohio State University,
Columbus, OH, 43210
[3] Electrical and Computer Engineering, The University of Utah,
Salt Lake City, UT, 84112

**Abstract:**

**We demonstrate self-powered β-Ga$_2$O$_3$ deep-UV metal-semiconductor-metal (MSM) photodetectors (PD) with 0.5% external quantum efficiency (EQE) at zero bias. 150 nm thick (-201)-oriented epitaxial β-Ga$_2$O$_3$-films were grown on c-plane sapphire using plasma-assisted MBE. Ni/Au and Ti/Au metal stacks were deposited as contacts to achieve asymmetric Schottky barrier heights in interdigitated finger architecture for realizing self-powered photodetectors. Current-voltage characteristics (photo and dark), time-dependent photocurrent and spectral response were studied and compared with conventional symmetric MSM PD with Ni/Au as the Schottky metal contact, fabricated on the same sample. The asymmetric, self-powered devices exhibited solar-blind nature and low dark current < 10 nA at 15 V with high photo-to-dark current ratio of ~ 10$^3$. The dark and photocurrents were asymmetric with respect to the applied bias and the responsivity in the forward bias was characterized by gain. The detectors (asymmetric-MSM) were found to exhibit a responsivity of 1.4 mA/W at 255 nm under zero-bias condition (corresponding to an EQE ~ 0.5 %), with a UV-to-Visible rejection ratio ~ 10$^2$ and ~10$^5$ at 0 V and 5 V respectively.**

[a)]Corresponding author email: anamika@iisc.ac.in, digbijoy@iisc.ac.in

β-Ga$_2$O$_3$ is thermodynamically the most stable phase among five other known phases (α, β, γ, δ, and, ε) of Ga$_2$O$_3$ with a wide bandgap of 4.6 eV [1,2], which makes it an attractive candidate for deep-UV detectors[3–9] and power transistors[10–15]. β-Ga$_2$O$_3$ also offers economic advantage as conventional crystal growth techniques such edge-defined film-fed growth (EFG)[16,17], float-zone[18–20] and Czochralski[21,22] methods can be employed towards enabling scalable and large-area single crystal wafers. Thin film growth of β-Ga$_2$O$_3$ on foreign substrates has also been widely reported using different growth techniques including Molecular beam epitaxy (MBE)[9,23–25], Metalorganic chemical vapour deposition(MOCVD)[26–28], mist CVD[29], radio frequency magnetron sputtering[30,31] and microwave irradiation approach[32] for different applications. β-Ga$_2$O$_3$-based deep-UV photodetectors with Schottky[33–35] and MSM[9,36–40] architectures on bulk and foreign substrate have been studied in the recent years. However, no report exists on β-Ga$_2$O$_3$ self-powered MSM detectors.

In this letter, we report on β-Ga$_2$O$_3$ self-powered lateral MSM PDs (asymmetric-MSM) with external quantum efficiency (EQE) of 0.5 % at zero bias. The current-voltage (I-V) characteristics, transient response and spectral response of asymmetric MSM (A-MSM) has been studied and compared with conventional symmetric Schottky MSM (S-MSM). This work reports the first zero-bias spectral responsivity for any type of UV detector based on epitaxial and planar β-Ga$_2$O$_3$.

Growth of β-Ga$_2$O$_3$ thin film on c-plane sapphire was carried out by plasma-assisted MBE equipped with a standard effusion cell for gallium and a Veeco Uni-bulb O$_2$ plasma plasma source. Following the substrate cleaning, the sapphire substrates were indium bonded to a silicon wafer and degassed at 400 °C for 1 hour in the buffer chamber before the actual growth run. Ga$_2$O$_3$ was grown for 3 hours at a substrate temperature of 700 C with a Gallium flux of 1.5 X 10$^{-8}$ and RF plasma power of 300 W. The β-Ga$_2$O$_3$ film was confirmed to be single phase (-201)-orientated with thickness of 150 nm from X-ray diffraction (XRD) and

X-ray reflectivity (XRR) measurements. The material characterization details have been reported earlier.[9]

Following standard lithographic process, Ti (20 nm)/Au (100 nm) stack was e-beam evaporated on the β-$Ga_2O_3$ films to form one side of the Schottky contact in the asymmetric MSM structure (Device: A-MSM). Post metal evaporation, a rapid thermal annealing (RTA) of Ti/Au metal stack was done at 470°C for 1 minute in nitrogen ambient resulted in improvement of the metal contacts in terms of current values. Subsequently, Ni (20 nm)/Au (100 nm) metal stack was evaporated for the other Schottky metal contact to obtain photodetectors of A-MSM geometry as shown in fig. 1(a). Conventional Schottky symmetric-MSM (Device: S-MSM) were also fabricated on the same sample with Ni (20 nm)/Au (100 nm) metal stack for both side of interdigitated architecture as shown in the fig. 1(b). Each device consisted of 36 interdigitated fingers with finger widths of 4 μm and spacing of 6 μm, resulting in an active area of 260 x 300 μm$^2$. The photo current of the devices was measured using Sciencetech, Inc. Quantum Efficiency (QE) setup consisting of a 150-W Xenon lamp, monochromator and Keithly-2450 source meter. For intensity/power values Xenon lamp was calibrated with a standard silicon photodiode attached to the QE set-up.

Figure 1(c) shows spectral response (SR) versus wavelength (λ) at an applied bias voltage of 5 V for the A-MSM device in forward and reverse bias. SR was calculated using the expression below:

$$SR_{Calculated} = \frac{I_{PHOTO} - I_{Dark}}{P.A} \quad (1)$$

where, $I_{PHOTO}$ is the photocurrent, $I_{DARK}$ is the dark current, P is the optical power density, and A is the active area (A = 300 x 260 μm$^2$). For A-MSM, the measurements in forward bias were carried out with positive potential on Ni/Au contact. The detectors exhibited a cutoff in responsivity at 253-255 nm. The peak SR values for A-MSM detector were measured to be 9.4 A/W and 1.1 A/W at 5 V in the forward (FB) and the reverse bias (RB) regimes, respectively. High internal gain in A-MSM detector was observed in the forward bias region as evident from Fig. 1(c). The zero-bias spectral responsivity is shown in the inset to figure 1(c) and a peak responsivity value of 1.4 mA/W was obtained corresponding to an EQE ~ 0.5 %, indicating its self-powered nature. It is more than an order of magnitude higher than the zero-bias responsivity of 0.01 mA/W[33] reported for β-$Ga_2O_3$-nanowire Schottky detectors, which is the only report till date on zero-bias SR for any kind of $Ga_2O_3$ UV detector. The UV to visible rejection ratio of the devices in this study was estimated by dividing the responsivity at 255 nm by that at 450 nm and was found to be ~$10^5$ at 5 V and ~$10^2$ at zero bias respectively, testifying the solar-blind nature of the photodetectors.

Further, voltage-dependent spectral response for both A-MSM and S-MSM was measured. Figure 2(a) shows SR versus wavelength at different applied biases in forward and reverse regions for A-MSM photodetectors. Figure 2 (b) shows the variation of peak SR (at 255 nm) with applied bias for both A-MSM and S-MSM PDs. With increasing applied biases in the forward/reverse region, the peak SR values (at 255 nm) were also found to be increasing while the overall SR in the forward bias regime was higher indicating a higher internal gain.

Figure 3(a) shows the variation of photo current and dark current with voltage (I-V) for the A-MSM detectors. The photo current was measured at an illumination of 255 nm while the bias on Ni/Au was swept with the Ti/Au contact grounded. Both the photocurrent and the dark I-V exhibited asymmetric natures with more than one order of rectification indicating asymmetric Schottky barrier heights for Ni/Au and Ti/Au contacts unlike the photo and dark currents of the S-MSM detectors (Fig. 3(b)) which exhibited symmetric behavior with applied bias. The asymmetric Schottky barrier heights for A-MSM detectors arise from the difference in the work functions of Ni and Ti metals used to form the interdigitated contacts on β-Ga$_2$O$_3$ thin film resulting in an asymmetric I-V. The photo currents at +15 V (FB) and -15 V (RB) for A-MSM detectors were thus measured to be 98 µA and 22 µA respectively while the corresponding dark currents were 100 nA and 1.7 nA. This can be contrasted with the corresponding photo (dark) current values of the S-MSM detectors (Fig. 3(b)). A-MSM devices were found to exhibit a photo to dark current ratio of ~ 10$^3$ at a bias of 5 V.

An approximate calculation of the Schottky barrier height (SBH) at the junction was done from the dark I-V characteristics, under the assumption that almost all the potential drop happens at the reverse-biased junction. SBH in forward and reverse junction was extracted using the modified Schottky equation below:[41]

$$I = I_0 \exp\left(\frac{eV}{nkT}\right)\left[1 - \exp\left(\frac{-eV}{kT}\right)\right] \quad (2(a))$$

$$and, I_0 = AA^*T^2 \exp\left(\frac{-e\phi_B}{kT}\right) \quad (2(b))$$

where, $I_0$ is the dark current, n is the ideality factor, e is electronic charge, V is applied bias, k is Boltzmann constant, T is temperature in Kelvin, I is the measured current, $\varphi_B$ is the barrier height, A is the area of the detector metal contact and A* is the Richardson constant. Based on a further simplification of the expression for MSM devices, the equation (2(a)) can be modified as below:[42]

$$I \exp\left(\frac{eV}{kT}\right) = I_0 \exp\left(\frac{eV}{nkT}\right) \quad (2(c))$$

Using the above equations 2(a), (b) and (c) and taking Richardson constant value ~ 41 A/cm$^2$K$^2$ for β-Ga$_2$O$_3$, SBHs for the A-MSM and S-MSM samples were extracted. The SBHs for A-MSM detector for Ti/Ga$_2$O$_3$ and Ni/Ga$_2$O$_3$ contacts were found to be 0.71 eV and 0.82 eV respectively whereas, the SBH for Ni/Ga$_2$O$_3$ contacts in S-MSM detector was found to be 0.82 eV. Thus, the lower Schottky barrier height at the Ti/β-Ga$_2$O$_3$ junction compared to the Ni/ β-Ga$_2$O$_3$ junction leads to a higher measured dark current in the forward bias region for A-MSM PD. Similar observations are reported in literature for GaN and AlGaN systems.[43–47]

Figure 4 (a) and (b) show the band diagram for A-MSM photodetectors in forward and reverse bias conditions (under illumination) respectively. When a negative bias is applied to the Ti/β-Ga$_2$O$_3$ junction, the photogenerated electrons drift towards the Ni contact and holes gets accumulated at the Ni/β-Ga$_2$O$_3$ junction. The junction thus gets positively charged. To maintain charge neutrality, more electrons from the Ti metal will flow towards the β-Ga$_2$O$_3$ leading to photo-induced lowering of the Schottky barrier height [9,48,49]. This results in a higher gain in the forward bias condition compared to reverse bias condition.

Figure 5 (a), (b) and (c) shows the transient current characteristics for reverse biased S-MSM (at -15 V), S-MSM (at -15 V) and A-MSM (0 V) detectors respectively. Rise times (10% - 90% of value) were estimated to be 2.5 s, 2.3 and 1.9 s while the fall times were found to be 0.4 s, 0.9 and 0.5 s for S-MSM (at -15 V), S-MSM (at -15 V) and A-MSM (0 V) detectors respectively. The on/off ratio for all samples was ~ 10$^3$. The slow response time can be attributed to hole trapping at the junction and in the bulk as reported in literature earlier.[9,48,49]

Table I shows a comparison of β-Ga$_2$O$_3$-based deep UV photodetectors in terms of spectral response, dark current and self-powered ability for various device architectures as reported in the literature. Although a zero-bias responsivity value of 0.01 mA/W[33] has been reported for Ga$_2$O$_3$-nanowire Schottky detectors, yet there has been no report on zero-bias responsivity for planar and epitaxial β-Ga$_2$O$_3$ UV photodetectors in either MSM or Schottky geometry other than this work.

In conclusion, we have demonstrated self-powered lateral MSM deep UV photodetectors based on β-Ga$_2$O$_3$. We have also carried out a comparative study of symmetric conventional MSM and asymmetric MSM in terms of dark current characteristics, voltage-dependent photoresponse, and time-dependent photoresponse. A spectral response of 1.4 mA/W at zero bias was demonstrated for asymmetric MSM photodetectors. Further, the spectral response analysis showed true solar-blind nature in conjunction with high photo-to-dark current ratio of ~ 10$^3$. This work reports zero-bias spectral responsivity for any type of epitaxial β-Ga$_2$O$_3$-based UV detector, and is expected to aid in the development of self-powered solar blind devices.

## Acknowledgement

This work was funded by Department of Science and Technology (DST) under its Water Technology Initiative (WTI), Grant No. DSTO1519 and Space technology cell (STC/ISRO). We would also like to thank Micro and Nano Characterization Facility (MNCF) and NNFC staff at CeNSE, IISc for their help and support in carrying out his work. Z.X. and S.R. acknowledge funding from the U.S. Office of Naval Research EXEDE MURI (Program Manager: Dr. Brian Bennett). This work was supported in part by The Ohio State University Materials Research Seed Grant Program, funded by the Center for Emergent Materials, an NSF-MRSEC, Grant No. DMR-1420451, the Center for Exploration of Novel Complex Materials, and the Institute for Materials Research.

# REFERENCES:

[1] T. Onuma, S. Fujioka, T. Yamaguchi, M. Higashiwaki, K. Sasaki, T. Masui, and T. Honda, Appl. Phys. Lett. **103**, 41910 (2013).

[2] H.H. Tippins, Phys. Rev. **140**, A316 (1965).

[3] S. Nakagomi, T. Momo, S. Takahashi, and Y. Kokubun, Appl. Phys. Lett. **103**, 72105 (2013).

[4] T. Oshima, T. Okuno, and S. Fujita, Jpn. J. Appl. Phys. **46**, 7217 (2007).

[5] T. Oshima, T. Okuno, N. Arai, N. Suzuki, S. Ohira, and S. Fujita, Appl. Phys. Express **1**, 11202 (2008).

[6] R. Suzuki, S. Nakagomi, Y. Kokubun, N. Arai, and S. Ohira, Appl. Phys. Lett. **94**, 222102 (2009).

[7] S. Kumar, A.S. Pratiyush, S.B. Dolmanan, S. Tripathy, R. Muralidharan, and D.N. Nath, Arxiv ID 1709.03692 (2017).

[8] T.-C. Wei, D. Tsai, P. Ravadgar, J.-J. Ke, M.-L. Tsai, D.-H. Lien, C.-Y. Huang, R.-H. Horng, and J.-H. He, IEEE J. Sel. Top. Quantum Electron. **20**, 3802006 (2014).

[9] A.S. Pratiyush, S. Krishnamoorthy, S.V. Solanke, Z. Xia, R. Muralidharan, S. Rajan, and D.N. Nath, Appl. Phys. Lett. **110**, 221107 (2017).

[10] H. Zhou, M. Si, S. Alghamdi, G. Qiu, L. Yang, and P.D. Ye, IEEE Electron Device Lett. **38**, 103 (2017).

[11] A.J. Green, K.D. Chabak, E.R. Heller, R.C. Fitch, M. Baldini, A. Fiedler, K. Irmscher, G. Wagner, Z. Galazka, S.E. Tetlak, A. Crespo, K. Leedy, and G.H. Jessen, IEEE Electron Device Lett. **37**, 902 (2016).

[12] M.H. Wong, K. Sasaki, A. Kuramata, S. Yamakoshi, and M. Higashiwaki, IEEE Electron Device Lett. **37**, 212 (2016).

[13] W.S. Hwang, A. Verma, H. Peelaers, V. Protasenko, S. Rouvimov, H.G. Xing, A. Seabaugh, W. Haensch, C. Van De Walle, Z. Galazka, M. Albrecht, R. Fornari, and D. Jeena, Appl. Phys. Lett. **104**, 203111 (2014).

[14] M. Higashiwaki, K. Sasaki, A. Kuramata, T. Masui, and S. Yamakoshi, Appl. Phys. Lett. **100**, 13504 (2012).

[15] K. Matsuzaki, H. Yanagi, T. Kamiya, H. Hiramatsu, K. Nomura, M. Hirano, K. Matsuzaki, and H. Honso, Appl. Phys. Lett. **88**, 92106 (2006).

[16] H. Aida, K. Nishiguchi, H. Takeda, N. Aota, K. Sunakawa, and Y. Yaguchi, Jpn. J. Appl. Phys. **47**, 8506 (2008).

[17] T. Oishi, Y. Koga, K. Harada, and M. Kasu, Appl. Phys. Express **8**, 31101 (2015).

[18] Y. Tomm, J.M. Ko, A. Yoshikawa, and T. Fukuda, Sol. Energy Mater. Sol. Cells **66**, 369 (2001).

[19] E.G. Villora, K. Shimamura, Y. Yoshikawa, K. Aoki, and N. Ichinose, J. Cryst. Growth **270**, 420 (2004).

[20] S. Ohira, N. Suzuki, N. Arai, M. Tanaka, T. Sugawara, K. Nakajima, and T. Shishido, Thin Solid Films **516**, 5763 (2008).

[21] Z. Galazka, K. Irmscher, R. Uecker, R. Bertram, M. Pietsch, A. Kwasniewski, M. Naumann, T. Schulz, R. Schewski, D. Klimm, and M. Bickermann, J. Cryst. Growth **404**, 184 (2014).

[22] K. Irmscher, Z. Galazka, M. Pietsch, R. Uecker, and R. Fornari, J. Appl. Phys. **110**, 63720 (2011).

[23] T. Oshima, T. Okuno, and S. Fujita, Jpn. J. Appl. Phys. **46**, 7217 (2007).

[24] X.Z. Liu, P. Guo, T. Sheng, L.X. Qian, W.L. Zhang, and Y.R. Li, Opt. Mater. **51**, 203 (2016).

[25] D. Guo, Z. Wu, P. Li, Y. An, H. Liu, X. Guo, H. Yan, G. Wang, C. Sun, L. Li, and W. Tang, Opt. Mater. Express **4**, 1067 (2014).

[26] M.J. Tadjer, M.A. Mastro, N.A. Mahadik, M. Currie, V.D. Wheeler, J.A.F. Jr, J.D. Greenlee, J.K. Hite, K.D. Hobart, C.R.E. Jr, and F.J. Kub, J. Electron. Mater. Electron. Mater. **45**, 2031 (2016).

[27] W. Mi, J. Ma, Z. Zhu, C. Luan, Y. Lv, and H. Xiao, J. Cryst. Growth **354**, 93 (2012).

[28] V. Gottschalch, K. Mergenthaler, G. Wagner, J. Bauer, H. Paetzelt, C. Sturm, and U. Teschner, Phys. Status Solidi A **206**, 243 (2009).

[29] K. Kaneko, H. Ito, S. Lee, and S. Fujita, Phys. Status Solidi C **10**, 1596 (2013).

[30] Y. Wei, Y. Jinliang, W. Jiangyan, and Z. Liying, J. Semicond. **33**, 73003 (2012).

[31] Z. Wu, G. Bai, Q. Hu, D. Guo, C. Sun, L. Ji, M. Lei, L. Li, P. Li, J. Hao, and W. Tang, Appl. Phys. Lett. **106**, 171910 (2015).

[32] P. Jaiswal, U. Muazzam, A.S. Pratiyush, N. Mohan, S. Raghavan, R. Muralidharan, and D.N. Nath, Appl. Phys. Lett. **112**, 021105 (2018)

[33] X. Chen, K. Liu, Z. Zhang, C. Wang, B. Li, H. Zhao, D. Zhao, and D. Shen, Appl. Mater. Interfaces **8**, 4185 (2016).

[34] T. Oshima, T. Okuno, N. Arai, N. Suzuki, S. Ohira, and S. Fujita, Jpn. J. Appl. Phys. **1**, 11202 (2008).

[35] R. Suzuki, S. Nakagomi, Y. Kokubun, N. Arai, and S. Ohira, Appl. Phys. Lett. **94**, 222102 (2009).

[36] D.Y. Guo, Z.P. Wu, P.G. Li, Y.H. An, H. Liu, X.C. Guo, H. Yan, G.F. Wang, C.L. Sun, L.H. Li, and W.H. Tang, Opt. Mater. Sexpress **4**, 1067 (2014).

[37] D. Guo, Z. Wu, P. Li, Q. Wang, and M. Lei, RSC Adv. **5**, 12894 (2015).

[38] G.C. Hu, C.X. Shan, N. Zhang, M.M. Jiang, S.P. Wang, and D.Z. Shen, Opt. Express **23**, 13554 (2015).

[39] T. Oshima, T. Okuno, and S. Fujita, Jpn. J. Appl. Phys. **46**, 7217 (2007).

[40] W.Y. Weng, T.J. Hsueh, S. Chang, G.J. Huang, and S.C. Hung, IEEE Trans. Nanotechnol. **10**, 1047 (2011).
[41] E. H. Rhoderick and R. H. Williams, Metal–Semiconductor Contacts ~Oxford University Press, New York, 1988.
[42] S. Averine, Y.C. Chan, and Y.L. Lam, Appl. Phys. Lett. **77**, 274 (2000).
[43] M. Brendel, M. Helbling, A. Knigge, F. Brunner, and M. Weyers, Electron. Lett. **51**, 1598 (2015).
[44] M. Brendel, F. Brunner, and M. Weyers, J. Appl. Phys. **122**, 174501 (2017).
[45] X. Chen, K. Liu, Z. Zhang, C. Wang, B. Li, H. Zhao, D. Zhao, and D. Shen, Appl. Mater. Interfaces **8**, 4185 (2016).
[46] D. Li, X. Sun, H. Song, Z. Li, H. Jiang, Y. Chen, G. Miao, and B. Shen, Appl. Phys. Lett. **99**, 261102 (2011).
[47] X. Sun, D. Li, Z. Li, H. Song, H. Jiang, Y. Chen, G. Miao, and Z. Zhang, Scientific Reports. **5**, 16819 (2015).
[48] A.M. Armstrong, M.H. Crawford, A. Jayawardena, A. Ahyi, and S. Dhar, J. Appl. Phys. **119**, 103102 (2016).
[49] O. Katz, V. Garber, B. Meyler, G. Bahir, and J. Salzman, Appl. Phys. Lett. **79**, 1417 (2001).

**Figure and Tables**:

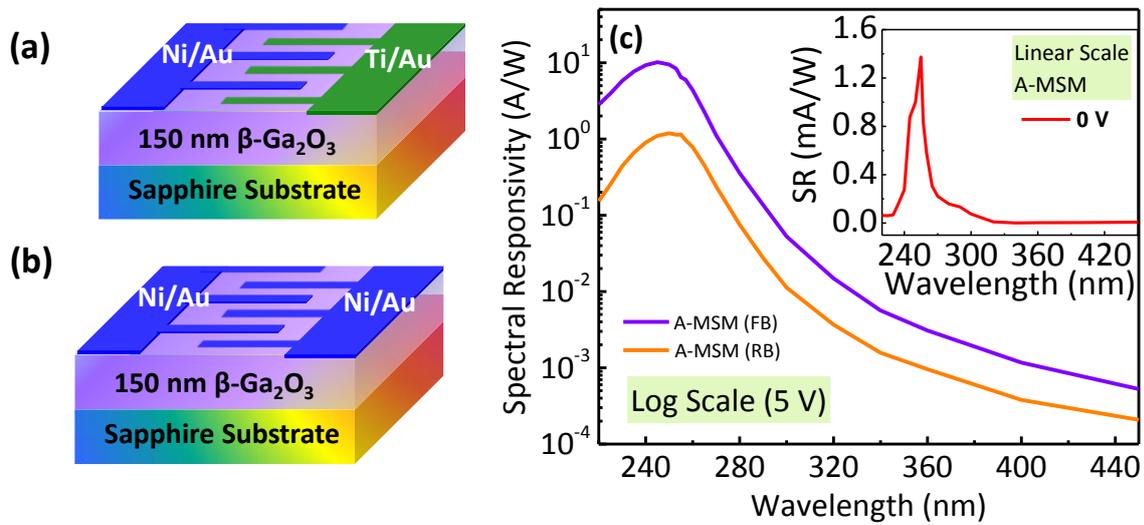

**Figure 1:** Schematic of **(a)** asymmetric-MSM (A-MSM) and **(b)** symmetric-MSM (S-MSM) device architecture respectively. **(c)** Spectral responsivity versus wavelength at an applied bias of 5 V for A-MSM detectors in forward and reverse bias condition. Inset of the figure 1(c) shows spectral response at zero bias for A-MSM photodetectors (EQE= 0.5%).

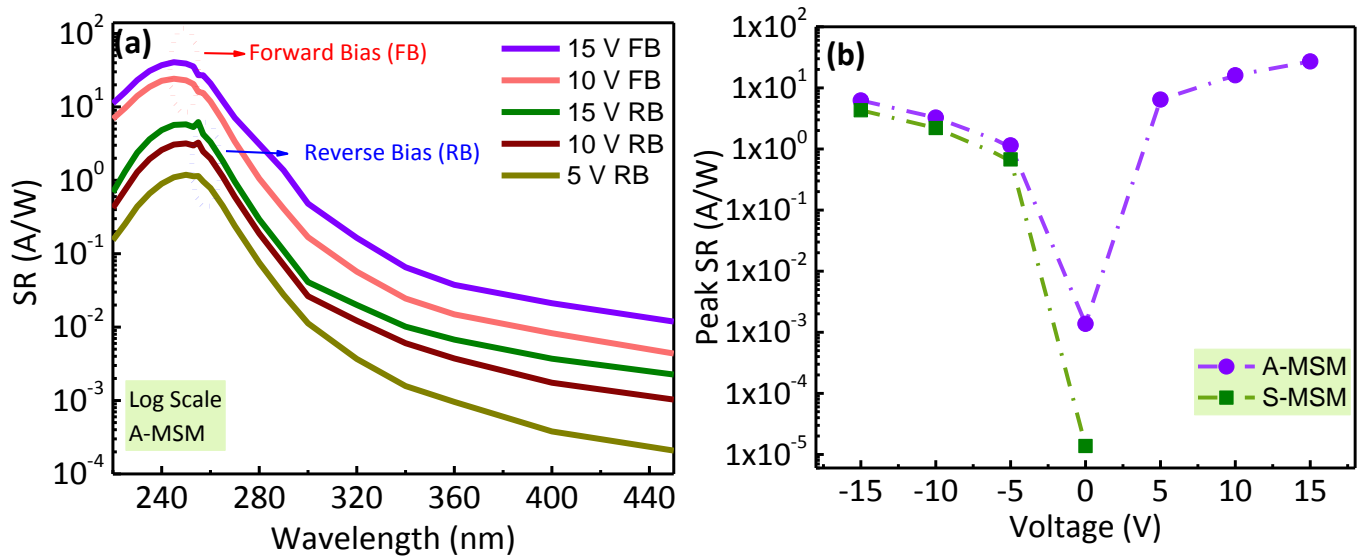

**Fig.2 (a)** Variation of spectral responsivity with the wavelength as a function of applied bias for A-MSM photodetector (Forward bias and Reverse bias) in log scale. **(b)** Peak response values (at 255 nm) versus bias voltages for A-MSM and S-MSM detectors.

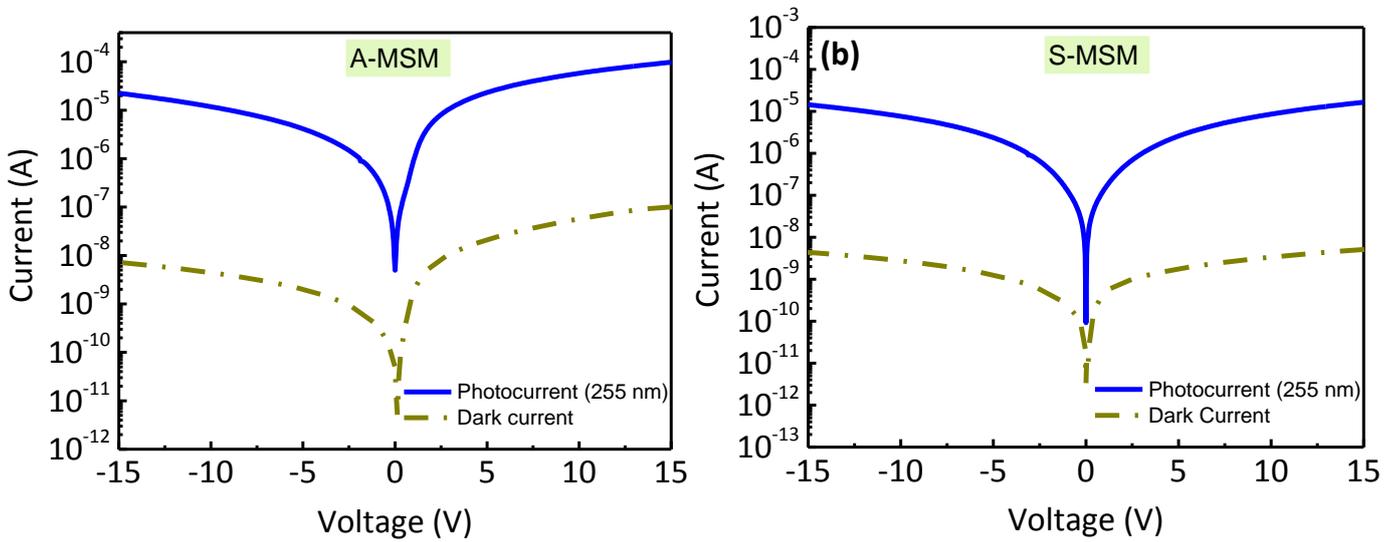

**Fig.3 (a)** Photo and dark I-V characteristics at room temperature for A-MSM detector (log scale). **(b)** Photo and dark I-V characteristics at room temperature for S-MSM detector (log scale). Illumination was done at 255 nm.

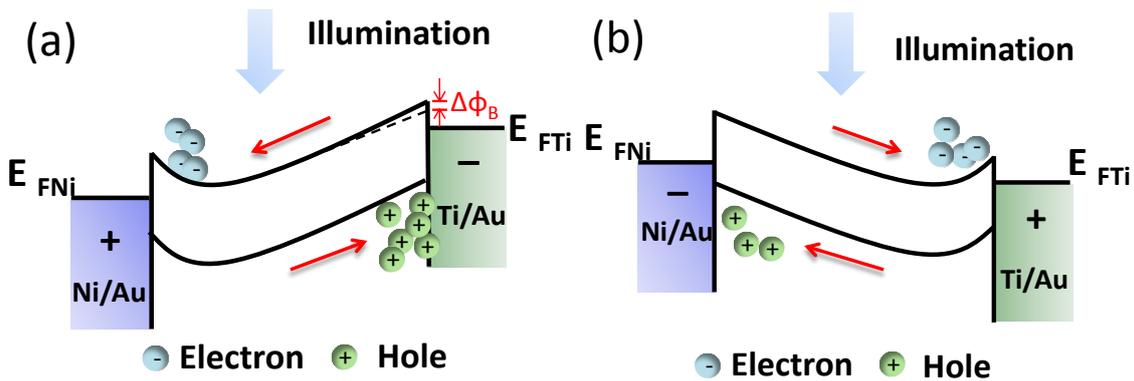

**Fig.4 (a)** Schematic of band diagram under UV illumination (255 nm) under **(a)** forward bias and **(b)** reverse bias condition for A-MSM photodetectors. ($E_{FNi}$ and $E_{FTi}$ denotes fermi level at different junction). **Fig. (a)** Photo induced lowering of barrier at metal-semiconductor interface by $\Delta\varphi_B$. (Dotted line represents new lowered barrier height after UV illumination in forward bias).

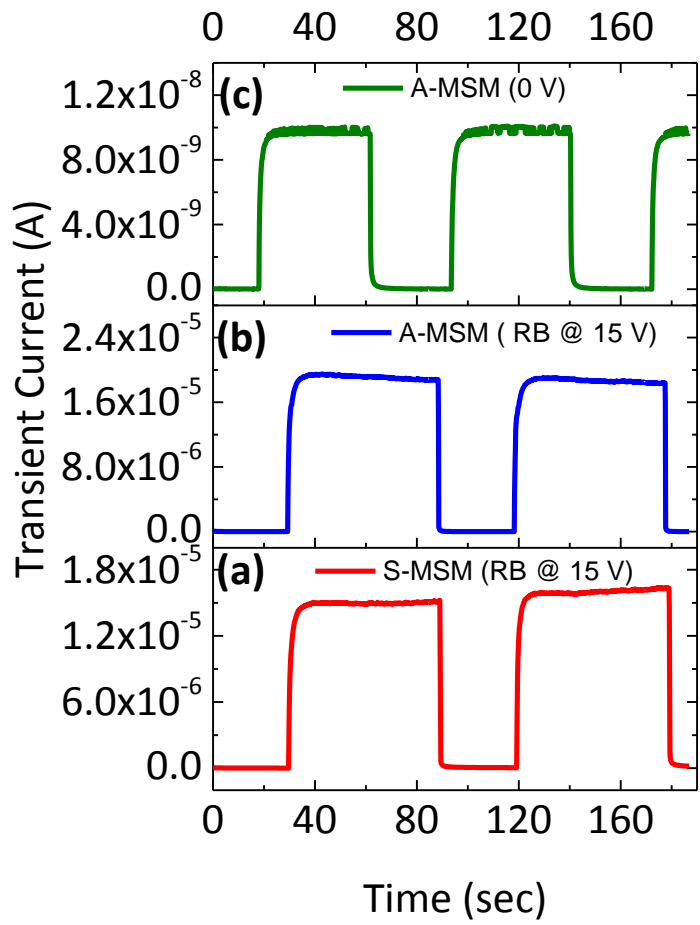

**Fig.5** Time-dependent photo-response (transient) under the illumination of 255 nm **(a)** reverse biased S-MSM at 15 V, **(b)** reverse biased A-MSM at 15 V and **(c)** A-MSM at zero bias respectively in linear scale.

**Tables**: (Double column)

TABLE I. List of β-$Ga_2O_3$-based deep UV detectors reported earlier for different architecture devices.

| Device Structure | Self-Powered | SR (Bias Voltage) | Dark current (Bias Voltage) | Reference |
|---|---|---|---|---|
| MSM | No | 37 mA/W (10 V) | 1200 pA (10 V) | 39 |
| MSM | No | 1.7 A/W (20 V) | 620 pA (20 V) | 38 |
| Schottky | Yes | 8.7 A/W (-10 V) | 10 nA (-10 V) | 34 |
| MSM | No | - | 128 nA (10 V) | 36 |
| Schottky | Yes | 1000 A/W (-3 V) | 0.1 nA (-3 V) | 35 |
| MSM | No | 0.37 mA/W (5 V) | 133 pA (5 V) | 40 |
| MSM | No | - | 5 nA (20 V) | 37 |
| MSM | No | 1.5 A/W (4 V) | 7 nA (20 V) | 9 |
| Schottky | Yes | 2.9 mA /W (-50 V) | 10 pA (-30 V) | 33 |
| A-MSM | Yes | **1 mA/W (0 V)** <br> 0.3 A/W (-5 V) | 7.3 nA (-15 V) | This Work |